\documentclass[preprints,article,accept,moreauthors,pdftex]{Definitions/mdpi}
\firstpage{1}
\pubvolume{1}
\issuenum{1}
\articlenumber{0}
\pubyear{2022}
\copyrightyear{2022}
\externaleditor{Academic Editor: {Naurang L. Saini}}
\datereceived{22 December 2021}
\dateaccepted{25 February 2022}
\datepublished{}
\hreflink{https://doi.org/} 

\Title{Band Structure of Organic-Ion-Intercalated (EMIM)$_x$FeSe Superconductor}

\TitleCitation{Band Structure of Organic-Ion-Intercalated (EMIM)$_x$FeSe Superconductor}

\Author{Lyudmila 
V. Begunovich $^{1}$\orcidA{} and {Maxim} M. Korshunov $^{1,2,}$*\orcidB{}}

\AuthorNames{Lyudmila V. Begunovich and  Maxim M. Korshunov}

\AuthorCitation{Begunovich, L.V.; Korshunov, M.M.}

\address{%
$^{1}$ \quad Siberian Federal University, Svobodny Prospect 79, 660041 Krasnoyarsk, Russia\\ 

$^{2}$ \quad Kirensky Institute of Physics, Federal Research Center KSC SB RAS, Akademgorodok, \linebreak 660036 Krasnoyarsk, Russia}

\corres{Correspondence: mkor@iph.krasn.ru}

\abstract{The band structure and the Fermi surface of the recently discovered superconductor (EMIM)$_x$FeSe are studied within the density functional theory in the generalized gradient approximation. We show that the bands near the Fermi level are formed primarily by Fe-$d$ orbitals. Although there is no direct contribution of EMIM orbitals to the near-Fermi level states, the presence of organic cations leads to a shift of the chemical potential. It results in the appearance of small electron pockets in the quasi-two-dimensional Fermi surface of (EMIM)$_x$FeSe.}

\keyword{band structure; DFT; Iron-based superconductors}

\begin{document}

\section{Introduction}
\begin{figure}[H]
\includegraphics[width=0.5\linewidth]{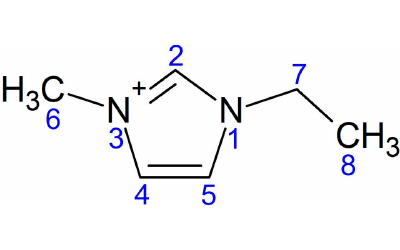}
\caption{Structural formula of EMIM cation. \label{fig:scatch}}
\end{figure}

Metal–organic compounds are a recent trend in functional materials design because of the combination of molecule flexibility in creating the framework and conducting, semiconducting, and topological features of the metal ions subsystem. For example, intensively studied tetraoxa[8]circulene~\cite{Yu2014,Baryshnikov2014,Kuklin2018} with integrated Li or Na ions are suggested to be a conductor and even a superconductor with the Ca ions~\cite{Begunovich2021}. Pure organic compounds are usually the low-temperature superconductors with the critical temperature $T_c$ of the order of 10~K, whereas metal–organic compounds demonstrate the higher $T_c$s. For example, potassium-doped $p$-terphenyl exhibits a superconducting transition temperature in the range from 7 to 123~K, depending on the doping level~\cite{WangR2017_1, WangR2017_2, WangR2017_3, Li2019}. Alternatively, one can use organic molecules as electron donors and structure stabilizers to control the features of a metallic system. It was initially suggested that the protonation via the ionic-liquid-gating method~\cite{Piatti2021} makes it possible to increase $T_c$ in iron selenide~\cite{Cui2019} due to the H$_y$-FeSe$_{0.93}$S$_{0.07}$ formation. The later study~\cite{Wang2021} uncovered the formation of an organic ion-intercalated phase (EMIM)$_x$FeSe, where EMIM stands for C$_6$H$_{11}$N$_2^+$ (see its structure in Figure~\ref{fig:scatch}). The EMIM cations were inserted into FeSe during the electrochemical process in the electrolytic cell with two platinum electrodes and EMIM-BF$_4$ as an ionic liquid. FeSe was placed on the cathode (negatively charged electrode) where the redox reaction takes place using the electrons transferred through an external circuit from the anode. It is not clear yet which of the chemical species gain electrons. Discovered superconductivity with $T_c$ about 44~K in this material brings up questions on the mechanism of Cooper pairing and on the role of EMIM molecules. Iron selenide belongs to a broad family of iron-based superconductors~\cite{y_kamihara_08,SadovskiiReview2008,IzyumovReview2008,IvanovskiiReview2008,JohnstonReview,PaglioneReview,LumsdenReview,StewartReview,HirschfeldKorshunov2011,Inosov2016} that also includes FeSe monolayer with $T_c$ above 80~K~\cite{FeSeTc,Zhang2015,GeFeSe100K,ZhaoFeSeLiOHFeSe,Sadovskii2016,Du2017,Liu2019,Jandke2019}.

\textls[-30]{To make a first step towards understanding the nature of superconductivity in (EMIM)$_x$FeSe,} here we calculate its band structure and Fermi surface using density functional theory (DFT). To place the EMIM molecules together with the FeSe lattice in the crystal structure, we construct a supercell corresponding to (EMIM)$_2$Fe$_{18}$Se$_{18}$. Bands near the Fermi level originate from the Fe-$d$ orbitals and orbitals of EMIM contribute at energies from 1 to 1.5~eV above the Fermi level. Although band structures of (EMIM)$_2$Fe$_{18}$Se$_{18}$ and FeSe with the same crystal structure are similar, their Fermi surfaces are different. In particular, small electron pockets around $X$-point appears in (EMIM)$_2$Fe$_{18}$Se$_{18}$. Therefore, we argue that the main role of (EMIM)$_2$ is to shift the chemical potential that results in the transformation of the Fermi surface.
\vspace{-15pt}

\section{Computation Details and Crystal Structure}

DFT~\cite{p_hohenberg_64,wkohn65} calculations were performed using open-source package for Material eXplorer software package (OpenMX)~\cite{Boker2011} based on a linear combination of pseudoatomic orbital (PAO) method~\cite{Ozaki2003,Ozaki2004,Ozaki2005,Lejaeghere2016} and norm-conserving pseudopotentials~\cite{Bachelet1982,Kleinman1982,Blochl1990,Troullier1991,Morrison1993}. The cutoff energy value was equal to 150~Ry. The PAO basis set $s2p2d2f1$ for Fe, $s2p2d2$ for Se, N and C, and $s2$ for H were set to describe the structures. Cutoff radii of 6.0~a.u. for Fe and H, 7.0~a.u. for Se, 5.0~a.u. for N and C were used. The generalized gradient approximation (GGA) proposed by Perdew, Burk and Ernzerhof (PBE)~\cite{jperdew96} was applied to describe the exchange-correlation effects. Empirical D3 correction of \mbox{Grimme~\cite{Grimme2010,Grimme2011}} was included to describe weak van der Waals interactions between EMIM cations. The criteria for the total energy minimization and interatomic forces were set to $1 \cdot 10^{-6}$~Hartree and $1 \cdot 10^{-4}$~Hartree/Bhor, respectively. The first Brillouin zone (BZ) was sampled on a grid of $6 \times 6 \times 6$~$k$-points generated according to the Monkhorst–Pack method~\cite{Monkhorst1976}. Band structure calculations were carried out along the high symmetry directions in the BZ: $\Gamma(0,0,0)-X(0,1/2,0)-M(1/2,1/2,0)-\Gamma(0,0,0)-Z(0,0,1/2)-R(0,1/2,1/2)-A(1/2,1/2,1/2)-Z(0,0,1/2)$, $X(0,1/2,0)-R(0,1/2,1/2)$, $M(1/2,1/2,0)-A(1/2,1/2,1/2)$. Maximally localized Wannier functions (MLWFs) were obtained by the Marzari–Vanderbilt procedure as implemented in the OpenMX package~\cite{Mazari1997,Souza2001}. The criterion for the minimization of the gauge invariant part of the spread function was set to $1 \cdot 10^{-8}$~\AA$^2$. The hybrid minimization scheme (steepest-descent and conjugate-gradients methods) was used to minimize the spread functional. The criterion for the minimization was equal to $1 \cdot 10^{-8}$~\AA$^2$. The Visualization for Electronic and Structural Analysis (VESTA) software~\cite{Momma2011} was used to represent the atomic structures.

\begin{figure}[H]
\centering
\includegraphics[width=1.0\linewidth]{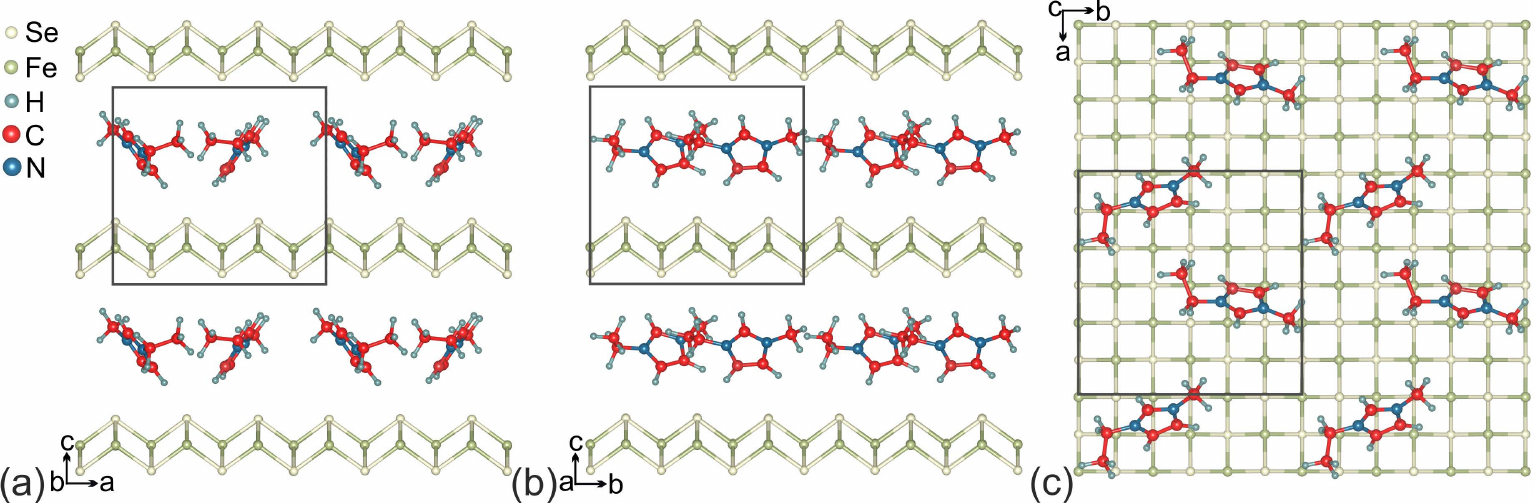}
\caption{Crystal structure of (EMIM)$_x$FeSe. The \textit{ab}-plane view (\textbf{c}) shows the EMIM arrangement within the layer. Unit cell is marked by a {rectangle}. \label{fig:cell}}
\end{figure}

It is known that the interaction between C--H bond and the $\pi$-system (C--H$\cdots \pi$ interaction) is observed in a large number of organic systems containing $\pi$-conjugated organic molecules~\cite{Tsuzuki2000}. According to Ref.~\cite{Yoshida2004}, the EMIM cations are also linked to each other by C--H$\cdots \pi$ interactions between one methyl carbon and the imidazolium ring of another cation. To provide a similar arrangement of EMIM cations between FeSe layers, we have chosen the $3 \times 3$~supercell of FeSe with two EMIM cations placed there. The formula for the resulting supercell is (EMIM)$_2$Fe$_{18}$Se$_{18}$; its structure is shown in Figure~\ref{fig:cell}.

The formation energy $E_{f}$ of (EMIM)$_2$Fe$_{18}$Se$_{18}$ was calculated using the following equation,
\begin{equation}
E_{f} = E_{(EMIM)_2Fe_{18}Se_{18}} - (E_{Fe_{18}Se_{18}} + 2E_{EMIM^+} + 2\varepsilon_F),
\end{equation}
where $E_{(EMIM)_2Fe_{18}Se_{18}}$ is the total energy of a (EMIM)$_2$Fe$_{18}$Se$_{18}$ system, $E_{Fe_{18}Se_{18}}$ and $E_{EMIM^+}$ are total energies of $3 \times 3$~supercell of bulk FeSe and EMIM cation, respectively, and $\varepsilon_F$ is 
the Fermi energy in the $3 \times 3$~supercell of a bulk FeSe.

Charge density difference induced by the interaction between the FeSe and EMIM cations is calculated as the difference in total charge densities between (EMIM)$_2$Fe$_{18}$Se$_{18}$ and superpositions of total charge densities of Fe$_{18}$Se$_{18}$ and (EMIM)$_2$, located at the same positions as in the (EMIM)$_2$Fe$_{18}$Se$_{18}$. These calculations were done for the same unit cell and with the same calculation conditions. Mulliken population analysis was used to estimate the charges on the atoms.

\section{Results and Discussion}

\begin{figure}[H]
\includegraphics[width=0.6\linewidth]{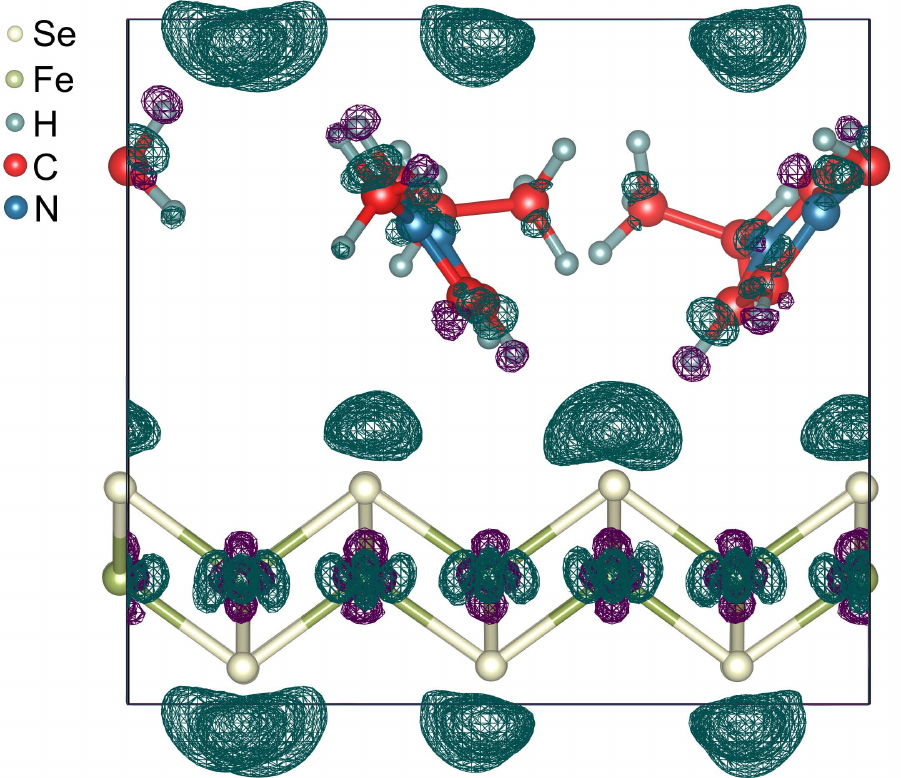}	
\caption{Charge density difference induced by the interaction between the FeSe and EMIM cations. Unit cell is marked by a rectangle. Dark teal and deep violet wireframe areas indicate electron-excess and electron-deficient regions, respectively. Isosurface level is $1.3 \cdot 10^{-3}$ ${a_0}^{-3}$, where $a_0$ is the \mbox{Bohr radius}. \label{fig:chg_dif}}
\end{figure}

The unit cell of (EMIM)$_2$Fe$_{18}$Se$_{18}$ are shown in  Figure~\ref{fig:cell}. The optimized lattice parameters are $a = b = 11.321$~\AA, $c = 10.450$~\AA. The length of $c$ parameter is in excellent agreement with the known experimental data~\cite{Wang2021}. EMIM cations are localized in the space between the nearest Se atoms of the FeSe layers. The smallest distance H--Se is 2.830~\AA. Localized EMIM cations make angles of $\sim$34$^{\circ}$ and $\sim$38$^{\circ}$ with the direction of the $c$-vector. The distance between the methyl carbon and the center of imidazolium ring is equal to 4.083~\AA, which is in the range of values for the C--H$\cdots \pi$ interactions~\cite{Tsuzuki2000}. The structural parameters of EMIM cations are presented in Table~\ref{tab:EMIM} and are in good agreement with the previous experimental and theoretical results for EMIM halide ionic liquids~\cite{WangY2005}. The Fe--Se bond lengths are slightly different and range from 2.325~\AA~to 2.338~\AA. These bond lengths are less than those for the unintercalated bulk FeSe~\cite{KumarR2010} and the bond-length difference ranges from $-2.6$ to $-3.1$\%. To examine the thermodynamic stability of (EMIM)$_2$Fe$_{18}$Se$_{18}$ the formation energy was calculated with the corresponding value of $-0.36$~eV per unit cell ($-0.02$~eV per FeSe formula unit). Negative values confirm that the formation of (EMIM)$_2$Fe$_{18}$Se$_{18}$ is energetically feasible. Analysis revealed the positive charge on EMIM cations ($+0.93~\bar e$ per one EMIM cation) and the negative charge on FeSe layers ($\sim$$-0.10~\bar e$ per FeSe formula unit). This means that the FeSe snatches electrons during an electrochemical reaction and becomes an anion. The snatched electrons are distributed between the atoms of FeSe so that the number of added electrons on one Fe atom and one Se atom is equal to $\sim$$0.02~\bar e$ and $\sim$$0.08~\bar e$, respectively. The total charge of the system is equal to zero. Thus, the EMIM cations intercalation through the electrochemical process allows one to perform an electronic doping of FeSe. In Figure~\ref{fig:chg_dif} we show that the electrons snatched by Se atoms are localized closer to EMIM cations (dark teal wireframe areas). This asymmetric electron density distribution results in a redistribution of the electron density in EMIM cations and on iron ions. A small number of electrons move from H atoms located near FeSe layer to C atoms. Further redistribution occurs by a chain mechanism.

\begin{table}[H]
\caption{Structural parameters of two EMIM cations located in FeSe. The atomic numbering scheme is shown in Figure~\ref{fig:scatch}. \label{tab:EMIM}}
\begin{adjustwidth}{-\extralength}{0cm}
\newcolumntype{C}{>{\centering\arraybackslash}X}
\begin{tabularx}{\fulllength}{CCCCCCCCC}
\toprule

\multicolumn{8}{c}{\textbf{Bond Length, \AA}} \\
N1--C2 & N1--C5 & N1--C7 & C2--N3 & N3--C4 & N3--C6 & C4--C5 & C7--C8 \\ \midrule
1.346 & 1.389 & 1.474 & 1.346 & 1.388 & 1.463 & 1.365 & 1.520 \\
1.346 & 1.389 & 1.475 & 1.347 & 1.387 & 1.461 & 1.364 & 1.519 \\ \midrule
\multicolumn{8}{c}{\textbf{Bond Angles in Degrees}} \\
N1--C2--N3 & C2--N3--C4 & N3--C4--C5 & C2--N3--C6 & C4--C5--N1 & C2--N1--C5 & C2--N1--C7 & N1--C7--C8 \\ \midrule
108.11 &108.82 & 107.18 & 125.16 & 106.92 & 108.96 & 125.56 & 112.12 \\
108.01 & 108.88 & 107.21 & 125.61 & 106.92 & 108.97 & 125.40 & 111.96 \\
	\bottomrule
		\end{tabularx}
	\end{adjustwidth}
\end{table}
\vspace{-6pt}

\begin{figure}[H]
\includegraphics[width=1.0\linewidth]{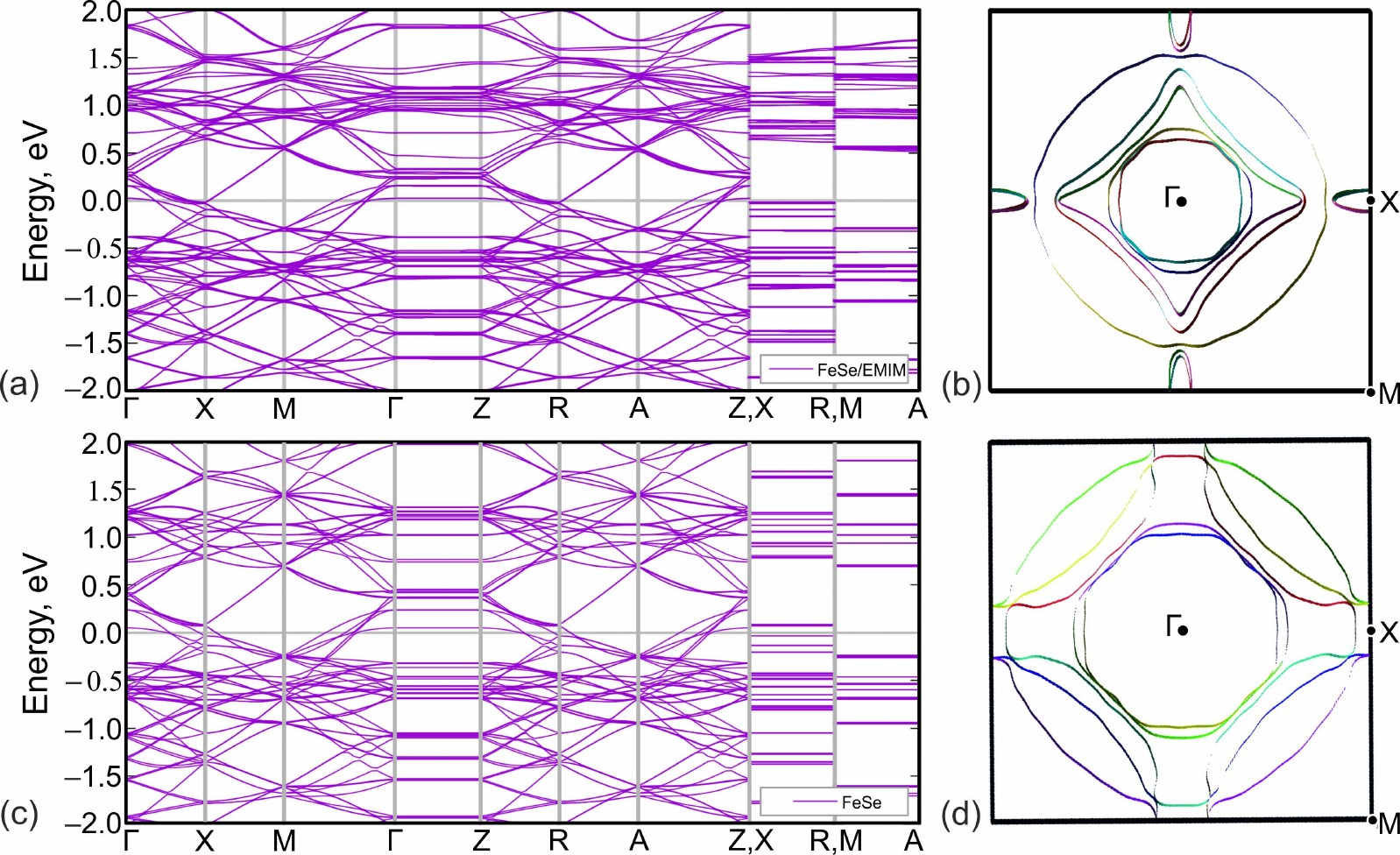}
\caption{Comparison of the DFT-calculated band structure (\textbf{a},\textbf{c}) and top views of Fermi surfaces (\textbf{b},\textbf{d}) of (EMIM)$_2$Fe$_{18}$Se$_{18}$ (\textbf{a},\textbf{b}) and of FeSe with the similar crystal structure (\textbf{c},\textbf{d}). The Fermi level corresponds to zero in panels (\textbf{a},\textbf{c}).  
\label{fig:comparison}}
\end{figure}

\begin{figure}[H]
	\includegraphics[width=0.7\linewidth]{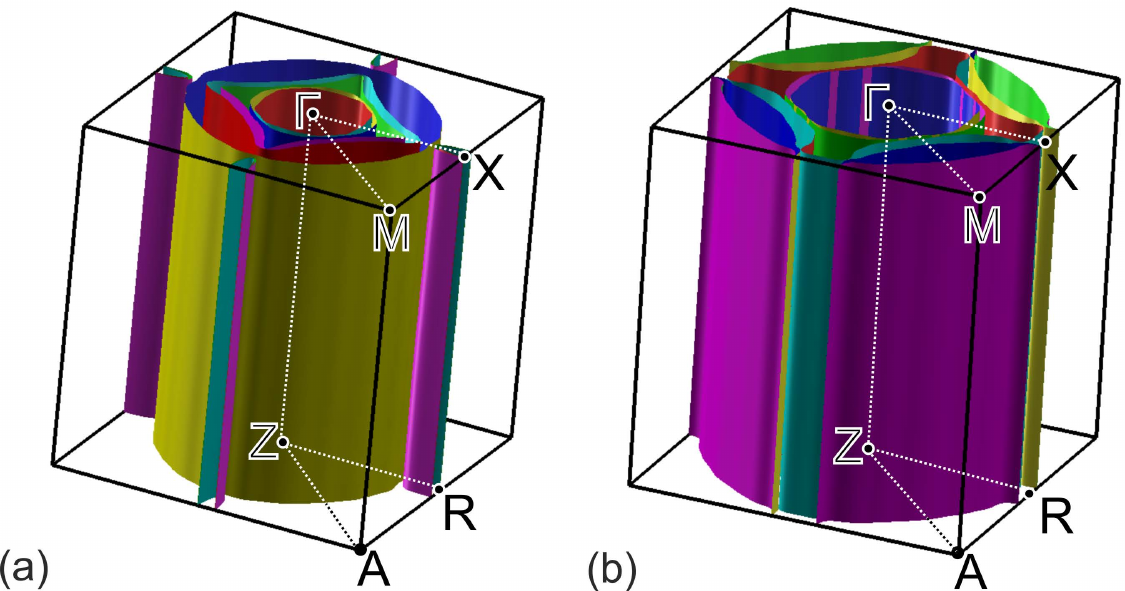}
	\caption{Perspective views of Fermi surfaces for (EMIM)$_2$Fe$_{18}$Se$_{18}$ (\textbf{a}) and FeSe with the similar crystal structure (\textbf{b}).
		\label{fig:fermi_surfaces}}
\end{figure}

The calculated band structure of (EMIM)$_2$Fe$_{18}$Se$_{18}$ is shown in Figure~\ref{fig:comparison}a. Bands cross the Fermi level in $\Gamma - X$, $M - \Gamma$, $Z - R$, and $A - Z$ intervals. Lack of dispersion of near-Fermi level bands in $\Gamma - Z$, $X - R$, and $M - A$ directions leads to a quasi-two-dimensional character of the Fermi surface, see Figure~\ref{fig:fermi_surfaces}a. According to Figure~\ref{fig:comparison}a,b, latter consists of two hole pockets around $\Gamma$-point, two almost degenerate small electron pockets around $X$-point, and three large Fermi surface sheets in-between these two points.

\vspace{-6pt}

\begin{figure}[H]
(a)\includegraphics[width=0.7\linewidth]{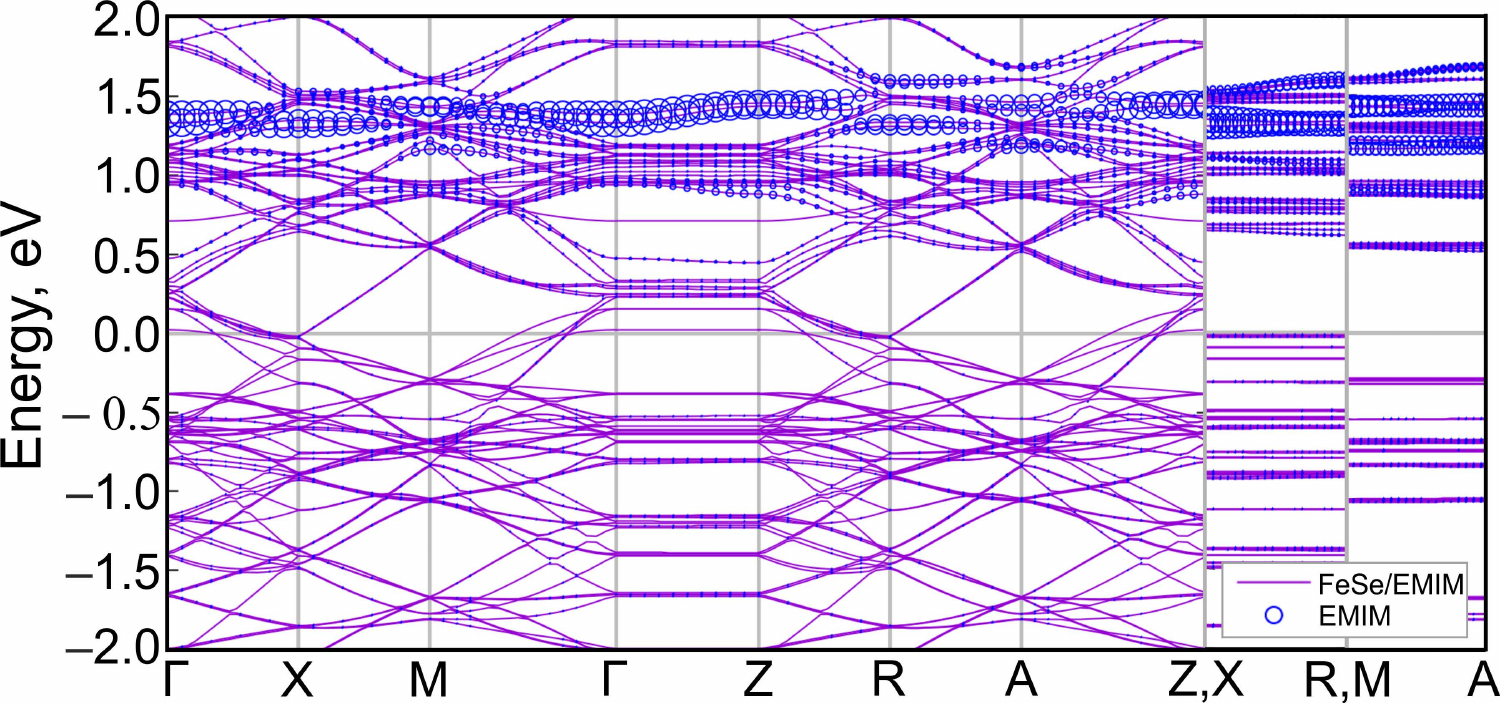}
(b)\includegraphics[width=0.7\linewidth]{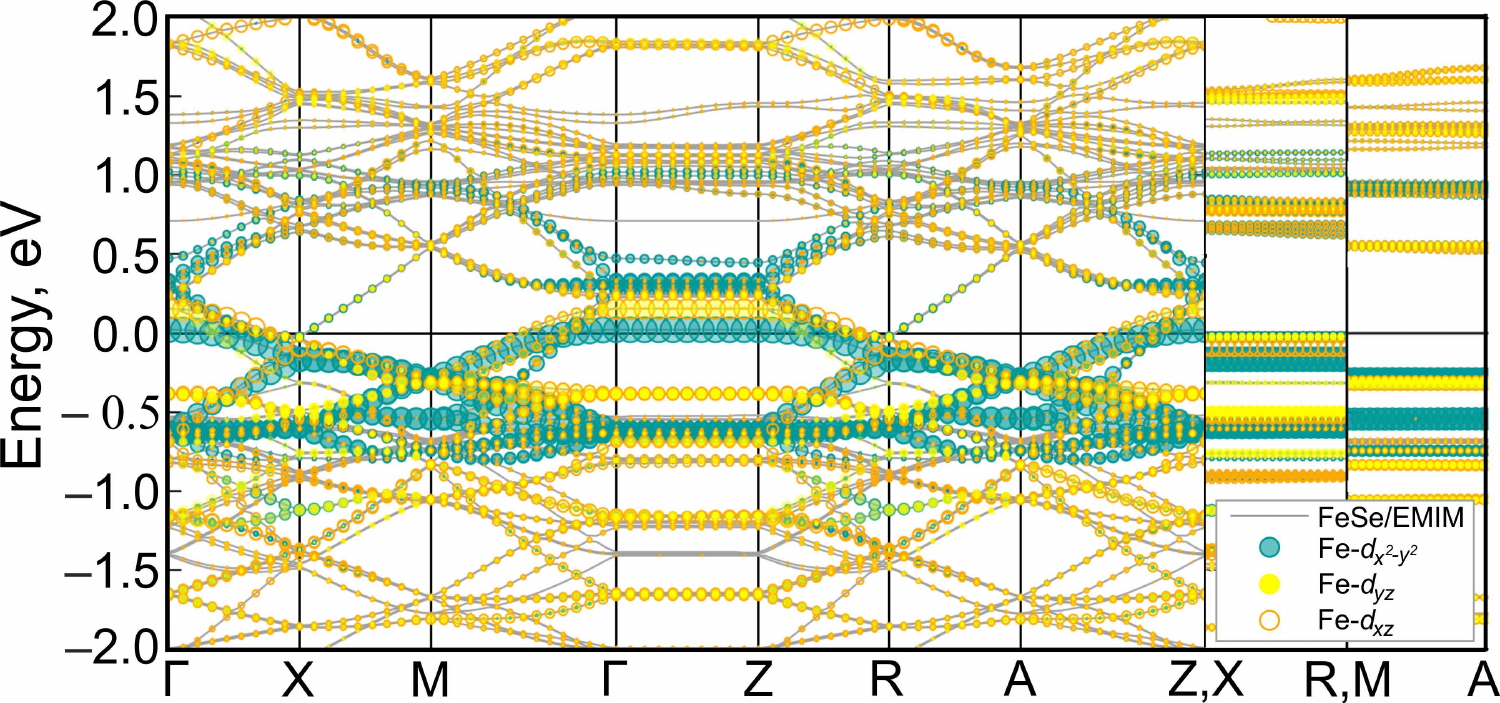}
(c)\includegraphics[width=0.7\linewidth]{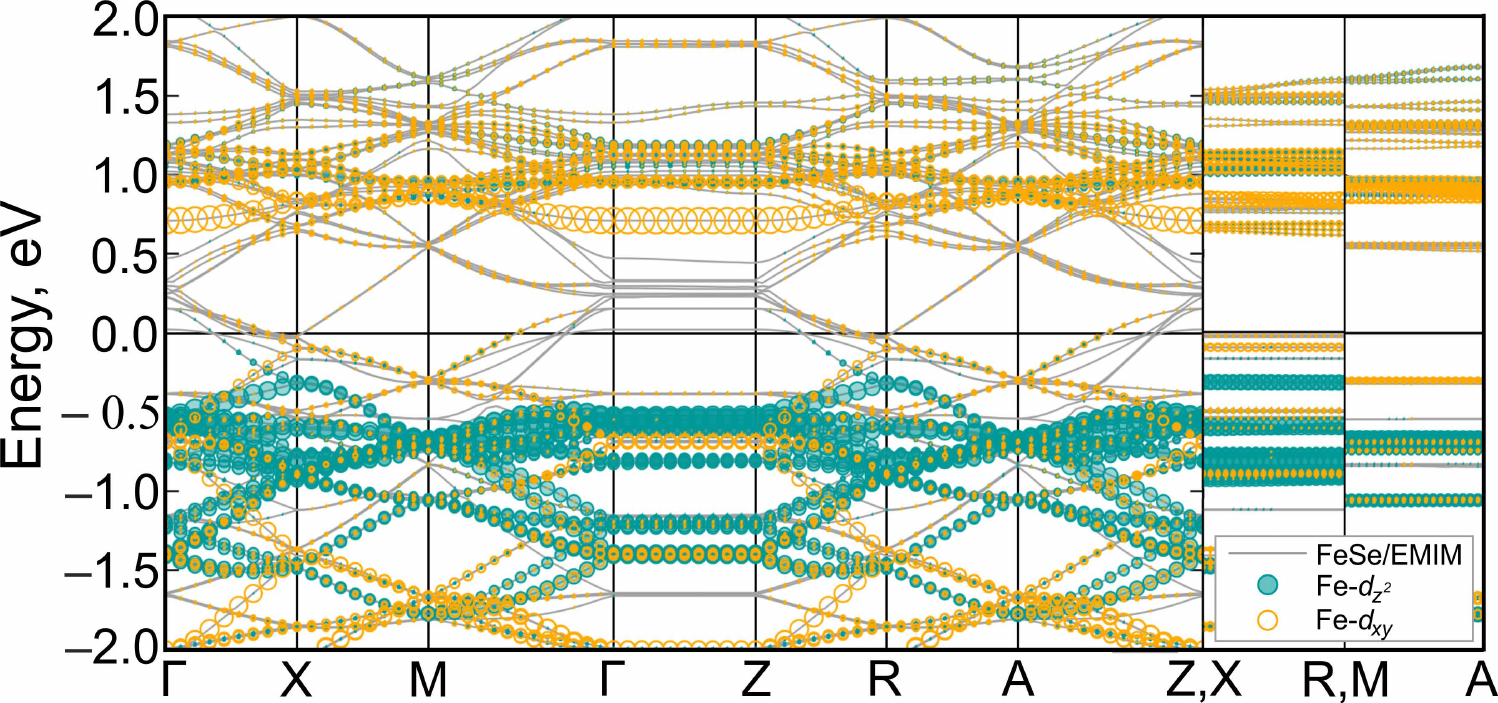}
\caption{DFT-calculated band structure of (EMIM)$_2$Fe$_{18}$Se$_{18}$ where the orbital contributions of EMIM (\textbf{a}), Fe-$d_{x^2-y^2}$, Fe-$d_{xz}$, Fe-$d_{yz}$ (\textbf{b}), Fe-$d_{xy}$, Fe-$d_{3z^2-r^2}$ (\textbf{c}) are shown by circles with sizes proportional to the corresponding weights. For visual clarity, the sizes are multiplied by 0.02 for EMIM and by 0.014 for all Fe-$d$ orbitals except for Fe-$d_{yz}$ that is multiplied by 0.012.
The Fermi level corresponds to zero. \label{fig:fatbands}}
\end{figure}

\begin{figure}[H]
\includegraphics[width=0.6\linewidth]{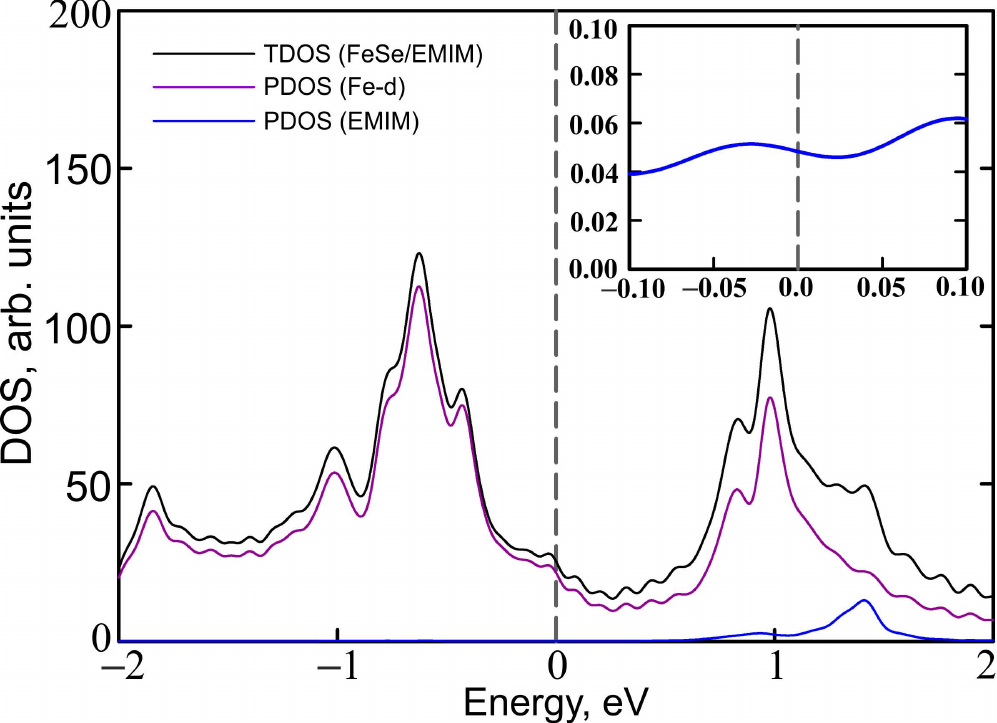}
\caption{Total density of states of (EMIM)$_2$Fe$_{18}$Se$_{18}$ (TDOS, black solid curves) and partial density of states (PDOS) of Fe 3$d$ orbitals (violet curves) and EMIM cations (blue curves). Artificial broadening is taken to be 0.05~eV. PDOS of EMIM cations in the vicinity of the Fermi level is shown in the inset. The Fermi level corresponds to zero.
\label{fig:dos}}
\end{figure}

The contribution of iron and EMIM orbitals to the band structure is shown in Figure~\ref{fig:fatbands}. Apparently, iron $d_{x^2-y^2}$, $d_{xz}$, and $d_{yz}$ orbitals contribute to the bands near the Fermi level (panel b), while most of the $d_{xy}$ and $d_{3z^2-r^2}$ orbital weight are located above and below zero (panel c). Partial density of states (PDOS) of iron $d$ orbitals form the most of DOS at the Fermi level (Figure~\ref{fig:dos}) with corresponding value of PDOS equal to 21.279~arb.units. Orbitals of EMIM (panel a) form separate bands around 1.5~eV and make a little contribution to the other bands located well above the Fermi level. At the Fermi level its contribution is very small (see the insert in Figure~\ref{fig:dos}). The PDOS of EMIM equal to 0.047~arb.units. that significantly smaller then PDOS of iron $d$ states. Therefore, there is no direct effect of EMIM on low-energy physics.

To analyze the overall role of organic cations, we compared the band structures of (EMIM)$_2$Fe$_{18}$Se$_{18}$, Figure~\ref{fig:comparison}a, and FeSe with the same crystal structure as (EMIM)$_2$Fe$_{18}$Se$_{18}$ complex but with the removed (EMIM)$_2$ cations, Figure~\ref{fig:comparison}b. Band structures are generally quite similar.
On the other hand, there are small differences around the Fermi level. Since there is no direct contribution of EMIM orbitals to those energies, they are caused by the electronic doping of FeSe layers in (EMIM)$_2$Fe$_{18}$Se$_{18}$.
Electron doping leads to a shift of the chemical potential, as a result of which more bands cross the Fermi level. All these lead to the change of the Fermi surface topology, compare Figures~\ref{fig:fermi_surfaces}a,b. Contrary to FeSe, Figure~\ref{fig:comparison}d, Fermi surface of (EMIM)$_2$Fe$_{18}$Se$_{18}$ contains small quasi-two-dimensional pockets around $X$-point, see Figure~\ref{fig:comparison}b.
Note the group of bands crossing the Fermi level near $X$-point and forming the aforementioned pocket in Figure~\ref{fig:comparison}a. The same group is above the Fermi level by about 100~meV in the FeSe system without (EMIM)$_2$ cations, see Figure~\ref{fig:comparison}c.

\begin{figure}[H]
\includegraphics[width=0.7\linewidth]{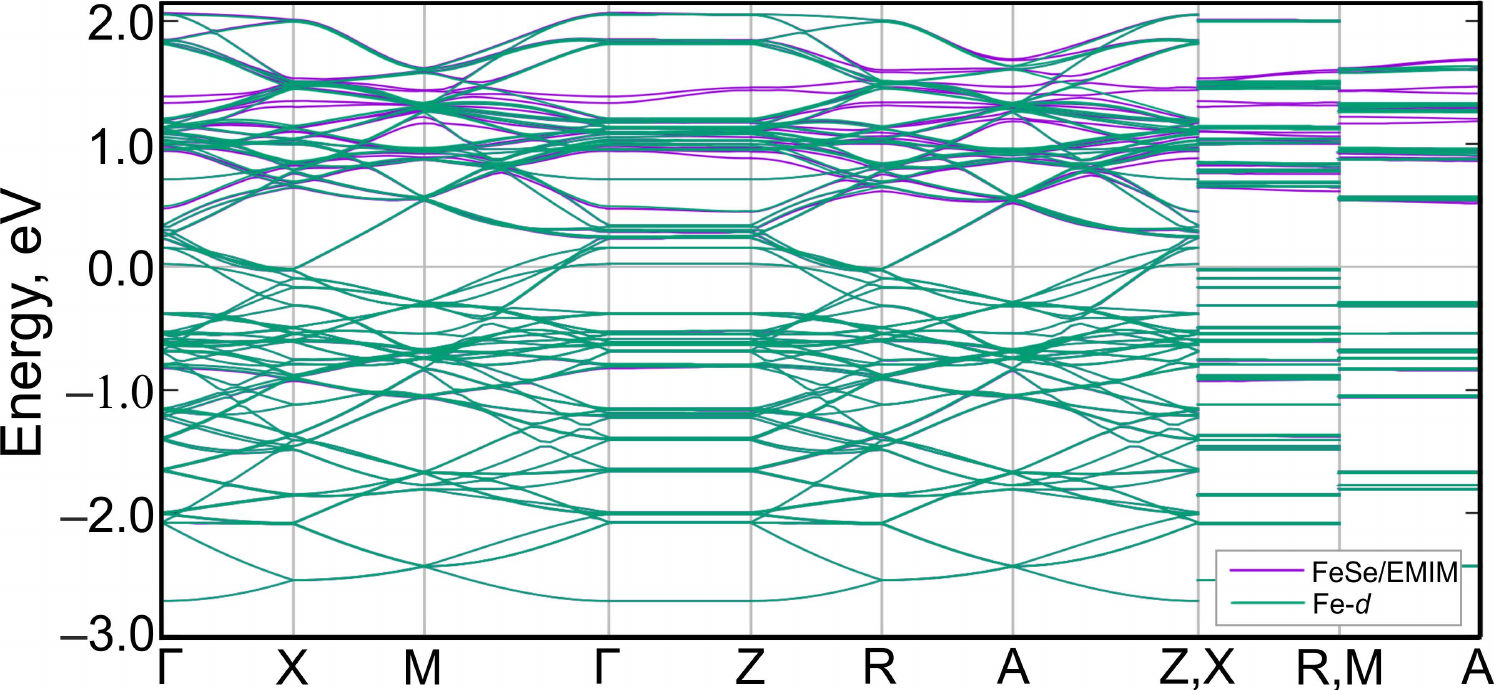}
\caption{Wannier-interpolated bands obtained from Fe-$d$ states (green curves) of (EMIM)$_2$Fe$_{18}$Se$_{18}$ and the original band structure (violet curves). Most of the interpolated bands overlap perfectly with the original bands. The Fermi level corresponds to zero.
\label{fig:WF}}
\end{figure}

Next, we derived the low-energy effective model from a set of Fe-3$d$-like Wannier functions. Wannier functions were generated from the Bloch states within the energy window from $-$3.00~eV to 2.15~eV. The constructed maximally localized Wannier functions are similar to classical $d$-orbitals and are centered on Fe atoms. Due to the size of the supercell, there are total of 90 Fe $d$-orbitals involved (five per iron, two irons in a unit cell, nine unit cells in the $3 \times 3$~supercell). The interpolated bands are shown in Figure~\ref{fig:WF} together with the original band structure. There is an excellent matching of interpolated and original bands in the vicinity of the Fermi level and in the valence band, which confirms that they are mainly $d$-like in character. Small mismatches appear above 0.5~eV in the conduction band, which is caused by the contribution of EMIM states. Some bands in the conduction band are absent, due to the small contributions of Fe-$d$ states to them. These bands are predominantly formed by EMIM orbitals. Thus, this effective model of the band structure based on all Fe-$d$ orbitals describes adequately the band structure of (EMIM)$_2$Fe$_{18}$Se$_{18}$ in the vicinity of the Fermi level.

\vspace{-6pt}

\section{Conclusions}

We have studied the electronic structure of (EMIM)$_x$FeSe using the state-of-art density functional theory in the generalized gradient approximation. In the absence of direct data on the coordinates of EMIM, we show that the reasonable position of it results in the (EMIM)$_2$Fe$_{18}$Se$_{18}$ unit cell. Fe-$d$ orbitals form the bands near the Fermi level similar to other Fe-based superconductors. Orbitals of EMIM do not affect the low-energy states directly; however, the presence of EMIM leads to the shift of the chemical potential that results in the transformation of the Fermi surface topology and appearance of small electron pockets around $X$-point in (EMIM)$_2$Fe$_{18}$Se$_{18}$ in contrast to FeSe with the similar crystal structure.

The appearance of the small electron pockets in EMIM-intercalated FeSe may play a crucial role in the formation of high-$T_c$ superconductivity. Spin-fluctuation theory of pairing~\cite{HirschfeldKorshunov2011,Kontani,Korshunov2014eng} predicts the sensitivity of the gap structure to the variation of the sizes of hole and electron pockets~\cite{MaitiKorshunovPRL2011,MaitiKorshunovPRB2011,MaitiKorshunov2012,ChubukovReview2012,FernandesReview2017}. Therefore, change of the Fermi surface topology upon EMIM intercalation may be the most important ingredient in the increase of $T_c$ reported in Ref.~\cite{Wang2021}.

Other important ingredients to the complete theory of superconductivity may come from the strong electronic correlations. Comparison of the LDA+DMFT (local density approximation + dynamical mean-field theory), a hallmark approach to include correlations, and the bare LDA revealed the weak influence of correlation effects on the electronic structure of the FeSe layer~\cite{Nekrasov2016}. On the other hand, experiments on monolayer FeSe have shown severe changes in the Fermi surface topology compared to DFT results~\cite{Liu2012}. To explain the discrepancy, strong electronic correlations, nematicity present in (EMIM)$_x$FeSe~\cite{Meng2021}, or something more exotic is needed. Here we made a first step towards making a theory that would include any of the mentioned mechanisms, i.e., we provide a DFT band structure on top of which the exotic mechanism can be built.

\vspace{6pt}



\authorcontributions{Conceptualization, M.M.K.; calculations, L.V.B.; writing, M.M.K. and L.V.B.; funding acquisition, L.V.B. All authors have read and agreed to the published version of the manuscript.}

\funding{This work was supported in part by Russian Science Foundation (Project 19-73-10015).
}

\institutionalreview{Not applicable}

\informedconsent{Not applicable}


\dataavailability{Not applicable}

\acknowledgments{L.V.B. would like to thank Information Technology Center, Novosibirsk State University, for providing the access to supercomputer facilities.}

\conflictsofinterest{The authors declare no conflict of interest.}

\begin{adjustwidth}{-\extralength}{0cm}

\reftitle{References}

\end{adjustwidth}
\end{document}